\documentclass{article}
\usepackage{booktabs} % Required for professional tables
\usepackage{cite}
\usepackage{caption}  % Required for proper table spacing
\usepackage{spconf, graphicx, amsmath, hyperref,setspace,enumitem, amssymb, bm}
\usepackage{subcaption}
\usepackage{multirow}
\usepackage[dvipsnames]{xcolor}
\usepackage{mathtools}
\usepackage[ruled]{algorithm2e}
\usepackage{hyperref}

\title{Interpretable Destination-Aware synthesizer Modulation Recovery }
\name{David Liu$^{1,2}$, Giulio Cengarle$^1$, David Cooper$^1$, Mark Vinton$^1$, Haici Yang$^1$ \thanks{Work done as part of an internship in Dolby. Thanks to Daniel Arteaga for providing insightful opinions during the project. LLMs were only used to assist with coding and grammar checking.}}
\address{$^1$Dolby Laboratories,  $^2$Boston University, USA}
\begin{document}
\ninept
\maketitle
\begin{abstract}
Synthesizer programming is a challenging task, particularly in configuring modulation: the time-varying control of parameters such as pitch, filter cutoff, and oscillator level by modulators such as Low-Frequency Oscillators (LFOs). Prior work has shown that modulation can be reconstructed from  the clean audio through curve matching, yet the recovered parameters are often untransferable to modern synthesizers. We present an interpretable destination-aware modulation and waveform recovery pipeline that mirrors how musicians often recreate sounds on a synthesizer. First, the system identifies the modulated destinations; then it recovers the LFO shape associated with each destination, and estimates the oscillator waveform to match the source timbre. The training is supported by a differentiable synthesizer that incorporates a noise oscillator, and more modulation options than previous work.  We carefully train the models for optimal perceptual quality, with extensive exploration of perceptual losses and adversarial training. Through objective and subjective evaluation, we show that predicting the destinations correctly is essential, our model excels in the modulation focused inverse synthesis tasks, and our Gammatone loss and CQT discriminator significantly outperform the traditionally-used MSS loss in improving perceptual quality. We provide audio samples from the subjective test\footnote{https://dolby.box.com/v/dest-aware-MUSHRA}.
% Gray box

% Re-synthesized audio is assessed with perceptual metrics including Multi-Resolution STFT (MSS), Mel-Frequency Cepstral Coefficients (MFCC) distance, and Frechet Audio Distance (FAD) with EnCodec. We show that our setup outperforms existing setups on objective metrics as well as human evaluations.
\end{abstract}

\section{Introduction}

Synthesizer programming is the process of configuring a synthesizer's parameters to get a desired audio output. When a music producer wants to use a sound they hear, they need to recreate the patch for the synthesizer, which can be challenging. 
% Even in the 1980s, widely used synthesizers such as the Yamaha DX7 offered rich but highly complex control spaces, leading many musicians to rely on factory presets rather than programming sounds from scratch. 
Modern synthesizers such as Serum\footnote{https://xferrecords.com/products/serum-2} and Vital\footnote{https://vital.audio/} are very expressive, exposing hundreds of parameters and modulation destinations, with many contemporary presets relying heavily on dynamic control signals such as LFOs. As a result, recreating a sound has become increasingly difficult, especially because synthesizer programming is fundamentally a many-to-one problem: multiple distinct parameter configurations can produce perceptually similar audio.

Recent advances in differentiable signal processing have significantly expanded the possibility of inverse synthesis programming. Early differentiable synthesizers \cite{engel2020ddsp, caspe2022ddx7, masuda2021synthesizer, hayes2021neural, zhao2020transferring} implemented classical synthesis techniques \cite{chowning1973synthesis, serra1990spectral} as differentiable modules, enabling end-to-end optimization of synthesis parameters directly from audio. While demonstrating capability in modeling real-world and synthetic sounds, these approaches faced limitations when applied to synthesis inversion for modern synthesizers.
% , as they lack the modularity commonly employed by contemporary musicians to create rich and varied timbres. 
To address this gap, subsequent work introduced more complex differentiable synthesizers incorporating modern modules \cite{uzrad2024diffmoog, turian2021one, bitra2025diffcraft} and a broader range of controllable parameters \cite{bruford2024synthesizer, masuda2023improving}. These systems support the inversion of multiple synthesizer parameters, predominantly in an end-to-end fashion via parameter regression losses \cite{barkan2019inversynth, masuda2023improving, uzrad2024diffmoog, yang2023white}, while others have turned to generative approaches \cite{esling2019flow, hayes2025audio} that trade interpretability for greater generative power.

\begin{figure}
    \centering
    \includegraphics[width=\columnwidth]{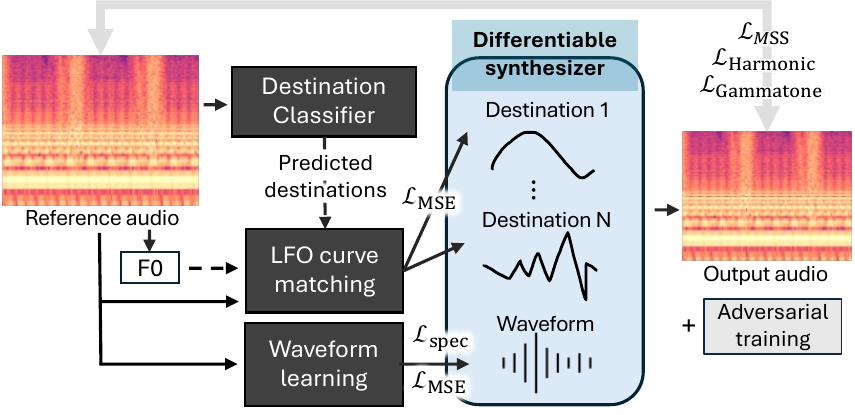}
    \caption{An interpretable inverse synthesis system featuring three major modules, which are connected by the differentiable synthesizer and trained with advanced techniques. Dotted lines indicate conditioning.}
    \vspace{-4mm}
    \label{fig:architecture}
\end{figure}
\label{sec:intro}

To further model the temporal expressiveness of sound, a line of work focuses on emulating the time-varying features of audio \cite{comunita2023modelling, wright2020neural}. In particular, ModSynth \cite{mitcheltree2025modulation} targets arbitrary modulation control, supporting three modulation types and providing an algorithm to extract arbitrary modulation curves (LFOs) built upon LFO-net \cite{mitcheltree2023modulation}, evaluated across white-, gray-, and black-box scenarios. Despite these advances, critical gaps remain. First, existing systems perform modulation extraction only on fixed destinations and clean audio, limiting their generalizability to practical use cases involving arbitrary destinations and more intricate sounds. Second, no existing method has attempted a holistic sound matching approach that simultaneously addresses LFO recovery across different destinations and waveform matching with perceptual losses. 

% Most closely related to the present work, Mitcheltree et al. proposed a modulation discovery framework that combines LFO extraction, constrained signal parameterizations, and DDSP sound matching to recover interpretable modulation curves from audio, explicitly investigating the trade-off between interpretability and reconstruction accuracy. Previously, LFO-Net, which extracts the LFO curves for modulated effects, such as phaser, flanger and chorus. Despite these advances, recovered parameters are not suitable for a general plugin.

% Generative approaches have further reframed the problem: Flow Synthesizer \cite{esling2019flow} used normalizing flows to model the distribution over synthesizer parameters, acknowledging the inherent ambiguity of the many-to-one mapping, while Hayes et al. \cite{hayes2025audio} treated inversion as a probabilistic inference problem across multiple synthesizer architectures. 

In this work, we aim at an interpretable inverse synthesis system that combines destination detection, modulation extraction, and waveform learning. As illustrated in Fig \ref{fig:architecture}, 
our approach mirrors the process a musician would follow when attempting to recreate a sound on a synthesizer: first identifying which modulation destinations are active, then characterizing the shape of the modulation applied to each destination, and finally matching the underlying oscillator waveform to reproduce the source timbre. 
% Active destinations are identified through training an Audio Spectrogram Transformer (AST) \cite{gong2021ast}-style classifier with multilabel classification, whose predictions then condition the downstream modulation recovery stage. 
Waveform recovery is a non-trivial step in this pipeline, as the modulations applied to a sound continuously alter its spectral shape over time, making it difficult to directly infer the underlying oscillator waveform from the audio alone.
We also attempt to tackle the inherent many-to-one issue by an extensive exploration of perceptual losses and including phase-invariant loss for waveform learning.

\section{Methods}

\subsection{Custom Synthesizer Design and Setup}
The synthesizer is a fully differentiable subtractive architecture 
% rendering audio at 2$\times$ oversampling (44,100 Hz) before decimating to 22,050 Hz by averaging pairs of samples, which reduces aliasing from nonlinear operations such as the tanh square wave approximation.  
supporting gradient flow from the audio output back through all the components, 
% the filters, oscillator, amplitude scaling, and modulation matrix, 
enabling end-to-end training against audio-domain losses. The signal path is as follows: oscillator and noise sources are amplitude-scaled and mixed, passed through a highpass then lowpass biquad filter in series, multiplied by an envelope, decimated, and finally RMS-normalized with a soft tanh limiter. 
% \noindent\textbf{Time-Varying Filters.}
Biquad filters are applied using \texttt{torchlpc}\footnote{https://github.com/DiffAPF/torchlpc} \cite{ycy2024diffapf} , which provides a parallelized recursive IIR filter implementation. 
% Without this, synthesizing a full clip at 176,400 oversampled samples per clip is prohibitively slow, making meaningful batched training infeasible. 
% The filter implementation is adapted from Mod Discovery, with the addition of batched audio rendering with full gradient support.

\noindent\textbf{Active Destinations.}
The modulation matrix routes LFO signals to five synthesis destinations, modulating parameters within the following ranges: 1) Oscillator level (-40 to +6 dBFS); 2) Oscillator coarse pitch ($\pm$12 semitones); 3) Noise level (-40 to +6 dBFS); 4) Highpass cutoff (0.2--8kHz); 5) Lowpass cutoff (0.2--18kHz).

\noindent\textbf{Oscillator/LFO curves. } The oscillator produces one-cycle waveform. The LFO curves are loops with certain period. Both waveform and LFO are sampled at 22kHz. All of the LFOs are absolute, except for pitch, which is relative to the base pitch. 

% \noindent\textbf{Data/LFO Lengths }
% The training data consists of samples with 1 second loops. A single loop is given to the model. The sample rate of the audio is $22050$ Hz. We have experimented with varying loop lengths and have found to be able to extract the period of one single loop well, given that the loops are identical. In future work, we will continue exploring the idea of varying loop lengths, including different loop lengths for different active destinations, but this is out of the scope for the current project.

% \haicicom{\textbf{missing} the lengths of oscillator and LFO curves, and sampling rate.}

% \vspace{-2mm}
\subsection{LFO and Waveform Recovery Pipeline}
\noindent\textbf{Step 1: Destination Classification.}
Given an input audio signal, the first step in this pipeline is to identify the destinations that are actively modulated. We train a multi-label classification model for this task, as it is common for more than one destination to be modulated simultaneously in a sound. The backbone model is expected to: 1) capture the necessary acoustic features of the input signal, and 2) model the signal's temporal context over timescales of up to several seconds in order to detect modulation patterns over time. We choose an AST-style \cite{gong2021ast} classifier for this task. AST is a purely attention-based model originally designed for audio classification. Its self-attention layers provide the ability to capture long-range global context.
% which has exceled in music and audio classification tasks . 
We adapt and train the AST model from scratch to have a smaller hidden dimension (128 vs. 768) and fewer number of layers (4 vs. 12). The number of parameters in this smaller model is around 800k, while acheiving similar performance on our task. The destination classification model is trained independently from the subsequent regression models, supervised by a binary cross-entropy loss.

% \haicicom{missing story}
% We train an Audio Spectrogram Transformer on the task of multilabel active destination classification. 
% \haicicom{Why using AST}
% Since this is independent from the curve matching models, we can train this separately. We train with torch's BCEWithLogitsLoss, and take destinations with their sigmoided values above 0.5 as active. 

% ADD CLASSIFICATION RESULTS HERE

% \section{Custom Synthesizer Setup}
% - Synthesizer has gradients 
% - uses torchlpc for the modulated filters for the speed
% \subsection{Time-varying filters}
% -- without the torch lpc thingy, synthesizing is very slow, to the point where it is not possible to have meaningful training
% -- take the filter implementation similar to mod discovery 
% -- added a way to render batches of audio with gradients
% \subsection{Active Destinations}
% -oscillator level
% -oscillator pitch
% -noise level
% -hp cutoff
% -lp cutoff

\noindent\textbf{Step 2: Destination-conditioned LFO Regression.}
% Rough idea draft
% WILL UPDATE THIS, WRITING DOWN IDEAS
% Adopting the SpectralCNN2D model from the modulation discovery paper, we embed the predicted destinations into the input spectrogram before the CNN layers rather than train a model for each destination. 
We activate zero to two simultaneous modulation destinations per sample during training. The regression model must isolate the contribution of the queried destination while suppressing interference from the concurrent modulation. 
% Training a separate model per destination is computationally prohibitive and does not generalize to unseen destinations. 
We adopt the SpectralCNN2D architecture from the previous work \cite{mitcheltree2025modulation} and condition it on the target destination by embedding a destination representation directly into the input spectrogram prior to the convolutional layers. This allows a single model to handle all destinations jointly. In addition to destination conditioning, we also use $f_0$ conditioning, in order to give context on the base frequency, specifically for relative pitch modulation. The $f_0$ conditioning is added to the spectrogram in the same way as the destination conditioning, but also added to the output projections for a stronger conditioning.
We directly apply the MSE loss for the LFO curves matching,  and process the entire spectrogram as one batch. We found that chunking the input into small windows and reducing input length doesn't improve the modeling accuracy. 

% We also experimented with a windowing approach, where we have 100ms windows from the spectrogram as input and then concatenate the results to form the output curve. This showed no difference from the no windowing approach. Our current method is to process the entire spectrogram.

As a post-processing step, we apply a multi-stage cleaning pipeline that first smooths the raw LFO signal using a sliding window average, then detects structurally meaningful keypoints by identifying peaks, troughs, and inflection points via first and second derivative analysis, and finally reconstructs a clean fitted curve through those keypoints using either linear interpolation or cubic splines depending on how much the segment deviates from linearity.

% We begin with the simplest setting: each sample contains a single active modulation destination with fixed base synthesizer parameters and a fixed MIDI pitch. Under this controlled condition, the destination-conditioned SpectralCNN2D accurately reconstructs LFO shapes across all four canonical waveforms. We further incorporate the waveform model, which produces faithful reconstructions of the source waveform, confirming that the proposed architecture is well-suited to the single-destination setting.

% \subsection{Multi-Destination Regression}

% \haicicom{Add loss}
% In this setting, the BiLSTM bottleneck proves insufficient, as the recurrent architecture lacks the capacity to capture the cross-temporal dependencies introduced by overlapping modulations. Replacing the BiLSTM with Transformer encoders and increasing the number of output channels gives a better reconstruction, especially with the waveforms, leading to better perceptual metrics.

% \haicicom{\textbf{Missing $f_0$ conditoin}}

\noindent\textbf{Step 3: Waveform Modeling.}
% - To get the correct timbre, we also try to model the base waveform of the oscillator.
% - We use the exact same model setup as the LFO recovery model without any conditioning
% - We fi
In addition to LFO recovery, we model the base oscillator waveform to capture the timbral characteristics of the synthesizer. 
Because the oscillators only produce short one-cycle waveforms, we treat the time-domain samples of the waveform as parameters, as in previous wavetable learning methods \cite{krekovic2021concept, shan2022differentiable}, and model the parameters through shape-matching.
We use the same architecture as the LFO recovery model, but without any destination or $f_0$ conditioning. 
% As shown in table \ref{tab:arch_comparison}, smaller models fail to reliably recover the waveform, 
We first experimented with a sub-million small model which fails to reliably recover the waveform, with a notable degradation in perceptual metrics, specifically the high frequency harmonic content. Increasing model capacity resolves this, suggesting that waveform recovery requires sufficient representational capacity to disentangle timbral features from the modulation activity present in the input spectrogram.
We also incorporate phase-invariant losses for the waveform by taking the MSE distance between the magnitudes of the FFTs ($\mathcal{L}_{\text{spec}}$). This ensures we do not penalize different phases, as their sounds are similar. 

\noindent\textbf{Pitch Complications.}
Pitch modulation introduces two distinct challenges that require targeted treatment. First, when pitch is actively modulated, the oscillator waveform is continuously stretched and compressed in the time domain, making it harder to recover the starting waveform.
% , which has been shown in previous work \cite{turian2020imsorrylossspectrallybased}. 
Second, the recovered LFO curves for pitch destinations tend to match the correct shape but are vertically offset from the ground truth shape.
To address the vertical offset, we shift the recovered pitch LFO curves to align their mean with the $f_0$ derived from CREPE\cite{kim2018crepe}, leaving the shape of the curve intact. Increasing the waveform model size also fixed the issue with the waveform recovery. This results in better pitch matching in the resynthesized audio without having to retrain a separate or larger model specialized for pitch modulation.

\subsection{Perceptual Quality Improvement}

% \subsubsection{Waveform Losses}
% \subsubsection{Perceptual Loss}

% \haici{In addition to the shape matching curve using, we apply perceptual losses to .... We first applied MSS loss following the previous works \cite{engel2020ddsp, caspe2022ddx7, mitcheltree2025modulation}, and observed reduced quality. Particularly, we noticed a tendency of pitch mismatching which has been studied in previous work \cite{}. Thus, we further explored more perceptual losses for this task.  }

% \haici{(Also need to reduce the lines for the following loss functions)}

We add perceptual losses to co-train LFO and waveform regression to improve the perceptual quality, and tackle the many-to-one issue. 
We first applied multi-scale spectral (MSS) loss following the previous work \cite{engel2020ddsp, caspe2022ddx7, mitcheltree2025modulation}, and observed reduced quality, particularly a tendency of pitch mismatching which has been studied before \cite{turian2020imsorrylossspectrallybased}. Thus, we further explored more perceptual treatment for this task, including harmonic loss, multiscale mel-spectrogram loss  and Gammatone loss, and adversarial training.

% The Multi-Scale Spectral (MSS) loss computes reconstruction error across STFT representations at multiple FFT resolutions, using the implementation provided by AURALOSS NEED TO CITE CHRISTIAN \cite{??}. At each scale,the loss combines a linear and log-magnitude L1 term over the magnitude spectrogram:

% \begin{equation}
%     \mathcal{L}_{\text{MSS}} = \frac{1}{|\mathcal{F}|} \sum_{k=1}^{|\mathcal{F}|}
%     \left(
%         \left\| |\mathbf{S}_k| - |\hat{\mathbf{S}}_k| \right\|_1
%         + \alpha \left\| \log|\mathbf{S}_k| - \log|\hat{\mathbf{S}}_k| \right\|_1
%     \right)
% \end{equation}

% where $|\mathbf{S}_k|$ and $|\hat{\mathbf{S}}_k|$ are the magnitude spectrograms of the target and reconstructed signals at scale $k$, and $\alpha$ balances the two terms. Aggregating across scales ensures the loss captures both fine temporal structure at small window sizes and spectral resolution at large window sizes.
% \subsubsubssection{Gammatone Loss}
\noindent\textbf{Harmonic loss} penalizes differences in the relative harmonic structure. For each signal, a differentiable $f_0$ is first estimated, upon which it constructs a harmonic vector. The loss is the mean L1 distance between the target and synthesized harmonic vectors.

\noindent\textbf{Multiscale mel-spectrogram loss} is adapted from MSS, computed across multiple mel scales and STFT sizes to capture perceptual differences at varying levels of time-frequency resolution.

\noindent\textbf{Gammatone loss} computes reconstruction error in the domain of a fixed, ERB-spaced Gammatone filterbank \cite{patterson1992complex}, approximating the frequency selectivity of the cochlea. A bank of $C = 64$ FIR filters is constructed from IIR Gammatone impulse responses  with center frequencies distributed on the ERB scale between 50~Hz and Nyquist. Each filter output is half-wave rectified and collapsed to a per-band RMS envelope over non-overlapping frames, yielding a representation $\mathbf{E} \in \mathbb{R}^{B \times C \times T}$.
The loss combines a linear and log-domain L1 term over the envelope, where $\lambda = 1.0$ and $\epsilon = 10^{-5}$.
\begin{equation*}
    \mathcal{L}_{\gamma} = 
    \left\| \mathbf{E} - \hat{\mathbf{E}} \right\|_1
    + \lambda \left\| \log(\mathbf{E} + \epsilon) - 
    \log(\hat{\mathbf{E}} + \epsilon) \right\|_1
\end{equation*}

\noindent\textbf{Adversarial Training.} 
We augment the reconstruction losses with an adversarial training objectives. We evaluate several discriminator architectures commonly used in neural audio synthesis: the Multi-Period Discriminator (MPD). Multi-Scale Discriminator (MSD), and Multi-Resolution Discriminator (MRD), and a Constant-Q Transform (CQT) discriminator \cite{kong2020hifi, lee2023bigvganuniversalneuralvocoder}, which operates on a logarithmically-spaced time-frequency representation better suited to musical signals.

\begin{table*}[h]

\vspace{-6.2pt}
\caption{Loss Function Configurations and Evaluation Metrics on Synthetic and Serum Data. The best metrics are bolded, and the second best are underlined.}
\centering
\label{tab:loss_configs}
\resizebox{\textwidth}{!}{%
\begin{tabular}{l ccc @{\hspace{1.5em}} ccc @{\hspace{1.5em}} ccc}
\toprule
 & \multicolumn{3}{c}{\textbf{Synthetic 2Dest}} & \multicolumn{3}{c}{\textbf{Synthetic 5Dest}} & \multicolumn{3}{c}{\textbf{\textit{Serum 2} 5Dest}}\\
\cmidrule(r){2-4} \cmidrule(l){5-7} \cmidrule(l){8-10}
\textbf{Configuration} & \textbf{MSS $\downarrow$} & \textbf{MFCC-L1 $\downarrow$} & \textbf{FAD $\downarrow$} & \textbf{MSS $\downarrow$} & \textbf{MFCC-L1 $\downarrow$} & \textbf{FAD $\downarrow$} & \textbf{MSS $\downarrow$} & \textbf{MFCC-L1 $\downarrow$} & \textbf{FAD $\downarrow$}\\
\midrule
MSE-only                                     & \underline{0.487} $\pm$ 0.330 & \underline{0.561} $\pm$ 0.348 & 0.898 & \underline{0.973} $\pm$ 0.757 & \underline{0.970} $\pm$ 0.635 & 0.899 & 2.445 $\pm$ 1.720 & \textbf{2.341} $\pm$ 0.685 & 49.386\\
MSE + MSS                               & 0.892 $\pm$ 0.505 & 0.822 $\pm$ 0.369 & 6.604 & 1.261 $\pm$ 0.696 & 1.131 $\pm$ 0.546 & 7.118  & 2.529 $\pm$ 1.693 & 2.392 $\pm$ 0.613 & 73.872\\
MSE + Gammatone                         & \textbf{0.469 $\pm$ 0.333} & \textbf{ 0.532 $\pm$ 0.349} & 1.865 & \textbf{0.961 $\pm$ 0.766}  & \textbf{0.952 $\pm$ 0.648} & 1.865 & \underline{2.443} $\pm$ 1.593 & \underline{2.350} $\pm$ 0.696 & \textbf{43.957}\\
MSE + Harmonic                          & 0.743 $\pm$ 0.356 & 0.798 $\pm$ 0.354 & 1.747 & 1.181 $\pm$ 0.703 & 1.169 $\pm$ 0.605 & 1.722 & 2.461 $\pm$ 1.696 & 2.400 $\pm$ 0.639 & 61.960\\
MSE + Mel-Spectrogram                   & 1.080 $\pm$ 0.552 & 0.953 $\pm$ 0.385 & 13.272 & 1.413 $\pm$ 0.687 & 1.235 $\pm$ 0.531 & 14.360 & 2.570 $\pm$ 1.953 & 2.356 $\pm$ 0.680 & 91.314\\
MSE + Disc. (CQT)                       & 0.520 $\pm$ 0.326 & 0.605 $\pm$ 0.347 & \textbf{0.591} & 1.000 $\pm$ 0.751 & 1.007 $\pm$ 0.631 & \textbf{0.591} & 2.445 $\pm$ 1.720 & 2.383 $\pm$ 0.750 & \underline{45.977}\\
MSE + Disc. (No CQT)                    & 0.528 $\pm$ 0.330 & 0.613 $\pm$ 0.347 & \underline{0.775} & 1.013 $\pm$ 0.758 & 1.017 $\pm$ 0.633 & \underline{0.700} & \textbf{2.440} $\pm$ 1.553 & 2.374 $\pm$ 0.650 & 46.915\\
MSE + Gammatone + CQT                   & 0.796 $\pm$ 0.442 & 0.790 $\pm$ 0.374 & 4.934 & 1.185 $\pm$ 0.674 & 1.124 $\pm$ 0.564
 & 4.200 & 2.522 $\pm$ 1.863 & 2.439 $\pm$ 0.630 & 75.467\\
MSE + ALL     & 1.116 $\pm$ 0.740 & 1.150 $\pm$ 0.634 & 8.683 & 1.399 $\pm$ 0.645  & 1.284 $\pm$ 0.532 & 9.069 & 2.695 $\pm$ 2.149 & 2.405 $\pm$ 0.649 & 85.865\\
No Conditioning (NC)                   & 1.034 $\pm$ 0.551 & 1.097 $\pm$ 0.571 & 2.554 & 1.466 $\pm$ 0.708 & 1.507 $\pm$ 0.725& 4.558 & 2.705 $\pm$ 1.608 & 2.577 $\pm$ 0.666 & 59.371\\
\midrule
% SpectralCNN2D (BiLSTM) + LFO + Waveform            & 0.644 $\pm$ 0.351 & 0.756 $\pm$ 0.374 & 2.005 & 1.051 $\pm$ 0.694 & 1.076 $\pm$ 0.681 & 2.095 & 3.000 $\pm$ 2.341 & 2.591 $\pm$ 0.655 & 75.998\\
% \midrule

Baseline (One Model Per Destination)         & 1.812 $\pm$ 0.728 & 1.901 $\pm$ 0.814 & 1.697 & 2.038 $\pm$ 0.717 & 2.106 $\pm$ 0.760 & 2.438 & 2.940 $\pm$ 0.941 & 2.792 $\pm$ 0.649 & 51.344\\
\bottomrule
\end{tabular}%
}
% \haicicom{what is spectralCNN2D(BiLSTM)}}
\end{table*}

\begin{table}[htbp]
  \centering
  \vspace{-6.2pt}
  \caption{Model architecture configurations and properties. Each model is trained with MSE + Gammatone losses. The models are evaluated on 2Dest. Scaling from \textit{Small} to \textit{Large} yields a substantial improvement. The waveform model reflects the same size as the LFO models (e.g. the small setup uses a small model for both LFO and waveforms). The baseline uses the same waveform model as the large setup.}
  
  \resizebox{\columnwidth}{!}{%
  \begin{tabular}{l c c c c c c}
    \toprule
    \textbf{Model} & \textbf{Params} & \textbf{Channels} & \textbf{Hidden} & \textbf{MSS} & \textbf{MFCC-L1} \\
    \midrule
    Small            & 0.8M & 64$\times$5 & 64 & 1.112 $\pm$ 0.537 & 1.030 $\pm$ 0.388\\
    Medium            & 2.7M & 64, 64, 128, 128, 128 & 128  & 1.052 $\pm$ 0.537 & 0.956 $\pm$ 0.387\\
    Large            & 7.6M & 64, 128, 128, 256, 256 & 256  & 0.469 $\pm$ 0.333 & 0.532 $\pm$ 0.349\\
    \midrule
    Baseline & 1.6M $\times$ 5 & 64, 64, 64, 64, 128 & 128 & 1.812 $\pm$ 0.728 & 1.901 $\pm$ 0.814\\
    \bottomrule
  \end{tabular}}
  \vspace{-4mm}
  \label{tab:arch_comparison}
\end{table}

\section{Experiment}
\subsection{Dataset and Training}
We generate synthetic datasets following the spline-based LFO shape procedure from Mitcheltree et al. \cite{mitcheltree2025modulation}. Each sample is synthesized using one of four canonical waveform shapes (sine, square, triangle, sawtooth) or a waveform drawn from the Serum \textit{analog} or \textit{spectral} wavetable banks. Task difficulty is controlled by varying synthesizer parameters specific to each experimental condition, as described in the following subsections.
The training data consists of samples with 1 second loops. A single loop is given to the model.  We experimented with varying loop lengths and found the model can extract the loop of different length well. The audio sample rate  is 22.05 kHz.
% #, given that the loops are identical. 
We use two main datasets: the 0-2 active destination dataset (2Dest) and the 0-5 active destination dataset (5Dest). 2Dest is used for both training and testing, and 5Dest is used for training the classifier as well as direct testing using models trained on 2Dest

% In future work, we will continue exploring the idea of varying loop lengths, including different loop lengths for different active destinations, but this is out of the scope for the current project.

We prepare both synthetic and \textit{Serum 2} data for testing. \textit{Serum 2} dataset is generated \cite{xfer_serum2_2025} through \texttt{Pedalboard} \cite{sobot_peter_2023_7817838} with the following configuration: one waveform oscillator, noise oscillator, filters, and LFO shapes generated in the same way as the synthetic datasets. The \textit{Serum 2} data under the same setup sound more refined compared to the synthetic datasets due to some unknown programming in the software; We treat it as a gray-box dataset.

% \subsection{Training}
The classifier and LFO/waveform models were trained independently. First, we train the destination classification model on 5Dest. Next, we train the LFO and waveform models concurrently on 2Dest using the MSE losses and optionally perceptual losses, while being conditioned on the ground truth $f_0$ and active destinations. During inference, $f_0$ is estimated using CREPE \cite{kim2018crepe} and destinations are predicted using the trained classifier before being used as conditioning for the LFO/waveform models.

% \haicicom{has the 0-to-2 and 0-to-5 properly explained anywhere}

All models were trained using the AdamW optimizer with a learning rate of 1e-4 on 1 A100 until convergence. A batch size of 64 was used for all models, except those trained with adversarial losses, which used a batch size of 32. 
For the loss weights, we found that $W_{\text{LFO}} = 1, W_{\text{Waveform}} = 0.1 $ works the best for the non-perceptual losses. For the perceptual losses, $W_{\text{Harmonic}} = W_{\text{Gammatone}} = W_{\text{Mel-Spectrogram}} = 1.0$ and $W_{\text{MSS}} = 0.5$.

\subsection{Baseline and Evaluation}
We designed a baseline and an ablation besides the losses variants to compare with our model. 
% \textit{Baseline-I} is designed as an easy baseline, where we remove the destination and $f_0$ conditioning from the proposed model, and keep other setup the same. 
The baseline has a competitive setup, in which a separate model is trained for each destination, akin to previous works \cite{mitcheltree2025modulation, mitcheltree2023modulation}, thus no destination classifier is technically needed. All other design choices, including $f_0$ conditioning, remain the same with the proposed model.
To make this baseline a fair comparison to our large models, we scaled the model size down so that the total number of parameters of every destination-specific models are similar to the single proposed model plus the classifier. We also trained it so that if a destination was not active, the target of a straight horizontal line at the correct value was given. This was done to prevent odd modulation curves when no modulation existed. 
% Changed this last line to fit 4 page limit

To test the function of the classifier itself, we design a separate ablation model, labeled as No-Conditioning (NC). The NC model keeps the same setup as the main model, but having both classification and $f_0$ conditioning removed, such that the LFO regression model has no information about which destination it is modeling.
Both the baseline and NC model are trained with MSE losses only. 

% Since we do not train a classifier for the baseline, every model is active during training and inference, which can explain some of the noise in the outputs. The waveform model is identical across all models.

% \subsection{Evaluation}
\noindent\textbf{Objective evaluation.}
% We evaluate our models on 2Dest as well as 5Dest. Our models are only trained on 2Dest, and is directly evaluated without finetuning on 5Dest. 
% We use MSS, MFCC and FAD-Encodec-24k as our metrics, and we also compare with 3 baselines the model with no conditioning trained on just LFO and Waveform losses, and the per-active-destination setup explained above.
% We also evaluate on a dataset generated with Serum 2 \cite{xfer_serum2_2025} through Pedalboard \cite{sobot_peter_2023_7817838}.
Following previous work \cite{mitcheltree2025modulation}, we use MSS, MFCC-L1 and FAD-Encodec-24k\footnote{https://github.com/microsoft/fadtk} as metrics to evaluate the synthesis quality of our models, from the perspective of spectral content reconstruction, human auditory perception, and perceptual distribution similarity, comparing all the variants with the baseline.
% We evaluate among different perceptual losses, and compared all the variant models with the baseline. Evaluations are conducted on both synthetic and \textit{Serum 2} data.

\noindent\textbf{Subjective evaluation.}
We designed a MUSHRA-style listening study, and modified the anchors from the standards catering to this specific task. For anchor 1, we used the same waveform as the ground truth, but no modulation on any destination. For anchor 2, we change the LFO shapes to be sine waves, for the same active destinations as the ground truth. We include five synthetic samples randomly selected from the test set with 1-5 active destination as well as 5 \textit{Serum 2} samples generated manually. We made sure the active destinations appeared in the test are balanced. Based on the objective scores and authors' listening,  the best three models (MSE losses only, MSE losses+Gammatone, MSE losses+Disc) and the basline are included in the test. 12 listening experts conducted the test.

% Using our model trained on 0, 1, 2 active destinations, we can accurately reconstruct the LFO shapes without retraining on more complex data. WRITE MORE HERE ABOUT THIS STUFF

\section{Results}
% \haiciadd{We can re-structure to 

% \begin{itemize}
%     \item Compare baseline and proposed models
%     \begin{itemize}
%         \item Baseline I: conditions are essential
%         \item Baseline II: Showed benefit of both conditioning; interpretability requires less data
%         \item if Table 2 is not discussed early, should include it.
%     \end{itemize}
%     \item Perceptual loss
%     \begin{itemize}
%         \item Basic analysis: Why MSE only is good; Why MSS, mel, harmonic may not work well; What possibly makes gammatone and CQT good; ``All'' doesn't work
%         \item Perceptual loss show advantages on out-of-domain data (in Serum2)
%     \end{itemize}
% \end{itemize}}

% \subsection{Classifier is important}
\subsection{Class Conditioning Analysis}
% From the results of Baseline I, conditioning on the active destination and $f_0$ are critical to a reasonable output.
% Without conditioning, every modulation shape is the same, regardless of the number of modulations. This is reflected on all of the perceptual metrics across all three datasets, specifically comparing to the MSEs only setup. 

\begin{figure}
    \centering
    \includegraphics[width=1\linewidth]{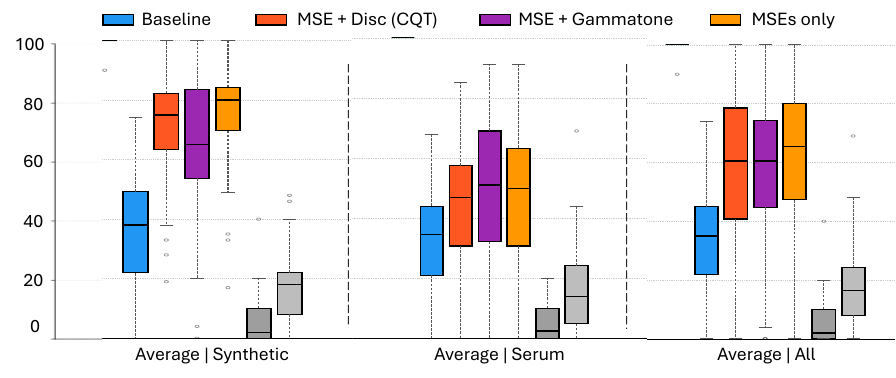}
    \caption{MUSHRA results. Gray boxes refer to the anchors, and the hidden reference are all located at 100, except for one outlier that is above 90. }
    \label{fig:placeholder}
    \vspace{-3mm}
\end{figure}

Both the objective evaluation in Tab.\ref{tab:loss_configs} and the MUSHRA scores in Fig.\ref{fig:placeholder} show that the baseline performs worse than all variants of our proposed model, demonstrating the importance of both classification and $f_0$ conditioning. 
% It's Despite having dedicated models for each destination and keeping all other model configurations identical, Th performance still falls significantly short of the conditioned models.
The baseline models were specifically trained to avoid cross-destination interference; however, we still observed that they occasionally confuse LFOs from other destinations with the current target destination, or produce non-zero curves when the destination is inactive, both of which contribute to the degraded quality. We attribute this to the fact that the baseline must simultaneously perform destination identification and curve matching. This design is less interpretable and makes the model more data-hungry and harder to optimize effectively. 

Interestingly, the NC model showed decent performance and outperformed baseline, which was unexpected, because the NC is not designed with the capability to correctly model every LFO curve. This result demonstrates that predicting correct destinations with a less accurate curve has more performance gain than better curve prediction with potential interference, highlighting the importance of the classifier.

% With every model being active, the amount of deviations increase with the number of destinations. This could potentially be fixed with a substantial increase in the dataset size. 
% Table \ref{tab:arch_comparison} further reinforces the importance of model capacity under the conditioning framework. 
% Scaling from Small (0.8M params, MSS = 1.112) to Large (7.6M params, MSS = 0.469) yields a substantial improvement in MSS, while Baseline II's ensemble of five models totals 8.0M parameters (1.6M × 5) yet achieves only MSS = 1.812.

\subsection{Perceptual Loss Analysis}

% - Talk about
% -- training with perceptual loss alone does not work
% --- most likely because it doesn't know which parameter affects which sound
% --- needs LFO/waveform loss
% -- perceptual loss doesn't make the metrics go down
% -- this might be due to the strict parameters of the synthesizer
% --- does not have absolute control over how the sound sounds
% --- Increasing the number of active destinations might let perceptual loss be a lot more useful

From Table \ref{tab:loss_configs} we find that the MSE + Gammatone setup performs best on MSS and MFCC-L1, followed very closely by the MSE-only setup, where no perceptual loss is used. The model trained with all discriminators,  including CQT achieves competitive scores, especially on the FAD score, which we attribute to discriminators' strength in modeling the data distribution. It is particularly interesting that, in the subjective test, the MSE losses have an edge overall. However, as it outperforms the other two perceptual loss configurations more evidently in the synthetic test set, the Gammatone loss is rated as the best system in the \textit{Serum 2} test set. 
Since ground truth active destinations are used during training, we alleviate the many-to-one problem in the training setup, allowing the direct MSE waveform losses to be highly influential while the perceptual losses help finetune the output. In the gray-box setup with \textit{Serum 2}, however, perceptual losses demonstrate their benefit.

MSS and mel-spectrogram losses suffer from the same failure modes, where high-frequency content dominates the loss and the resulting gradient is poorly correlated with perceptual distance between synthesizer patches. Harmonic loss can be very noisy due to samples with pitch modulation and incorrect $f_0$ predictions, particularly for samples containing noise or complex harmonics. Gammatone loss is the only perceptual loss to improve the metrics; it specifically models audio perception according to the human auditory system, which may explain its benefits over other perceptual losses.

Adding the CQT discriminator shows clear improvement over only using MPD, MSD, and MRD, which were designed for speech synthesis, not musical audio. The CQT's log-frequency axis with constant relative bandwidth means harmonic relationships appear as consistent patterns regardless of fundamental frequency, making it a more natural fit for discriminating synthesizer sounds and providing stronger gradient signal during training.

We found in our experiments that training with perceptual loss alone does not work, most likely because the model has no direct understanding of which synthesizer parameters affect which aspects of the sound, making it difficult to backpropagate meaningful gradients through the synthesis process.
% As a result, perceptual loss fails to drive the evaluation metrics down in any meaningful way. This limitation is likely compounded by the strict, constrained nature of the synthesizer itself. 
% Since the model does not have absolute control over how the output sound is shaped, the perceptual loss signal has little room to actually influence the result. 
Introducing an auxiliary LFO or waveform loss appears necessary to ground the optimization in more direct parameter-level supervision. It is also worth noting that increasing the number of active modulation destinations within the synthesizer could potentially make perceptual loss far more useful, as a more expressive synthesis space would give the model greater flexibility to respond to perceptual feedback.

% \section{Limitations}
% While our method can accurately reconstruct the LFO and waveform shapes of single oscillator audio, it fails to accurately reconstruct the sound of complex modulated audio from professional synthesizers such as Serum 2. One of the main causes for this are the filters: there are many different types of filters in Serum 2 while we only consider high pass and low pass. These filters greatly shape the audio; taking away the filtering results in similar audio to our training set. We also do not consider any effects, such as reverb, distortion or flanger, as modulating time varying effect parameters can vastly change the audio, making a much more complex problem, which we view is out of the scope of this project. 

% \section{Conclusion}
% We have shown that it is possible to extract LFO modulation shapes from specific destinations on noisy data with a single model. Currently, the model and synthesizer capacity does not match modern synthesizers, but the setup allows for easy additions to the active destinations. Future work will explore multiple oscillators, custom filters, and effect modulation. 

\section{Discussion}
% \section{Conclusion}
% We have shown that it is possible to extract LFO modulation shapes from specific destinations from noisy data with a single model. 
In this work, we presented a system that accurately identifies modulation destination, and reconstructs the LFO and waveform shapes of single oscillator audio, demonstrating that synthesizer reverse engineering is tractable even under noisy conditions. Moreover, the system delivers greatly improved synthesis quality with better perceptual treatment. However, the gap between our synthesizer model and professional synthesizers such as \textit{Serum 2} highlights clear directions for future work. The two main limitations of our current system are the restricted set of filter types and the absence of wavetable oscillators. Similarly, we do not model effects such as reverb, distortion, or flanger, as time-varying effect parameters introduce a level of complexity we consider out of scope for this work. Currently, the model and synthesizer capacity does not match modern synthesizers, but the architecture allows for easy extension to additional modulation destinations. Future work will explore multiple oscillators, custom filter types, and effect modulation, moving toward a system capable of reverse engineering the full complexity of professional synthesizer patches.

\bibliographystyle{IEEEbib}
\bibliography{strings,refs}

\end{document}